\documentclass[%
 amsmath,amssymb,
 aps,
]{revtex4-2}

\usepackage{booktabs}
\usepackage{multirow}
\usepackage{upgreek}
\usepackage{graphicx}
\usepackage{dcolumn}
\usepackage{bm}
\usepackage{dcolumn}
\usepackage[mathlines]{lineno}
\usepackage[utf8]{inputenc}
\usepackage[T1]{fontenc}
\usepackage{mathptmx}
\usepackage{etoolbox}
\usepackage{color}
\usepackage{ulem}

\begin{document}

\preprint{APS/123-QED}

\title{Frequency tuning of a squeezed vacuum state using interferometric enhanced Bragg diffraction effect}

\author{Qiqi Deng}
\author{Wenqi Li}
\author{Xueshi Guo}
\email[E-mail: ]{xueshiguo@tju.edu.cn}
\author{Xiaoying Li}%
\email[E-mail: ]{xiaoyingli@tju.edu.cn}
\affiliation{College of Precision Instrument and opto-Electronics Engineering, Key Laboratory of Opto-Electronics Information Technology, Ministry of Education, Tianjin University, Tianjin 300072, People’s Republic of China\\}

\begin{abstract}
%
%
We experimentally demonstrate the optical frequency tuning of a squeezed vacuum state generated from an optical parametric oscillator by using an acousto-optic modulator based bi-frequency interferometer.
The systematic efficiency of the frequency tuning device is $91\%$, which is only confined by the optical transmission efficiency of the acousto-optic modulators.
The amount of frequency tuning is 80 MHz, which is orders of magnitude larger than the line-width of the laser used to generate the squeezed state, and can in principle be further extended to GHz range. 
%
%
Our investigation shows the interferometric enhanced Bragg diffraction effect can be applied to a variety of other quantum optical states as well, and will serve as a handy tool for quantum network.

\end{abstract}

\date{\today}

\maketitle


\section{\label{sec:level1}INTRODUCTION}

Coherent interaction is of central importance in quantum optics, and interferometers are the fundamental blocks to many quantum information technologies.
To keep the coherence of quantum optical systems, it is vital to ensure the modes of all optical fields involved are well matched \cite{RevModPhys.92.035005, Tanzilli2005}.
The degree-of-freedom of optical field includes the polarization, spatial and spectral mode etc.. Different from polarization mode, which can be easily adjusted by using the combination of half wave plate and quarter wave plate, freely adjusting the frequency mode, which usually requires the involvement of nonlinear interaction, is not trivial. 
Compared to frequency conversion of classical light, quantum frequency conversion is more challenging. 
In particular for the continuous variable quantum state, the non-ideal converting efficiency will inevitably introduce vacuum noise, which degrades the performance of quantum information processing.
%
%
For the realization of frequency tuning with large amount, parametric process in $\chi^{(2)}$ or $\chi^{(3)}$ nonlinear optical medium is often exploited and high converting efficiency has been demonstrated \cite{{Huang1992PRL},{NicolasGisin2007OE},{Vollmer2014PRL}}.
%
However, a frequency tuning of a continuous variable quantum state in radio frequency (RF) range has not been demonstrated.
%

Optical parametric process is the workhorse for generating quantum states, whose coherence time is ultimately limited by the spectrum line-width of the pump laser.
%
For a practical quantum network, larger scale coherent interaction among independent quantum sources are indispensable.
At the current stage, the line-width of laser is sub-kilo Hertz or narrower and the coherent time is larger than 1 ms \cite{{2021Sub-kHz-Linewidth},{2014External}}, which is sufficient for most of the coherent operations. 
However, it is still not trivial to build two laser sources with the same frequency within the order of their spectral line-width, so quantum states generated from the same laser source are often used in order to effectively establish coherent operations to each other \cite{2013Ultra}. 
Alternatively, the central wavelengths of different pump lasers for state generation can be coincided by using optical phase locking loop \cite{{ol-39-10-2998},{seishu2019robust}}.
%
%
Both technical approaches have to access the laser that serves as the pump of the parametric process.
However, pump lasers in different locations of a quantum network are not always accessible.
In this case, 
acousto-optic modulators (AOMs) can in principle be used to shift the frequency of the quantum states directly. However, the non-ideal efficiency of individual AOM will inevitably introduce vacuum.
%
Therefore, A frequency tuning device with radio frequency (RF) tuning range, which can directly apply to quantum state with high efficiency is desirable.

Recently, our group successfully developed an AOM based bi-frequency interferometer (ABI), which can be used to handle quantum optical state with high systematic efficiency \cite{2022arXiv221000406L}. 
%
However, a real quantum state has not been used as a input state to illustrate that the ABI is particularly suitable for quantum technology.
Here we demonstrate a frequency tuning scheme by applying interferometric enhanced bragg diffraction effect of the ABI to a squeezed vacuum state, which is an optical state possesses quantum property without ambiguity. 
In our experiment, a squeezed vacuum state with -3.47 $\pm$ 0.02 dB squeezing rate is generated from an optical parametric oscillator (OPO) below threshold. 
After tuning the frequency of squeezed vacuum by the ABI,  
-1.98 $\pm$ 0.02 dB squeezing rate is directly measured by a fiber coupled homodyne detector.  
The amount of frequency tuning here is 80 MHz, which is already orders of magnitude larger then the line-width of the laser severs as the pump for nonlinear interaction.
Indeed, the frequency tuning range of AOM is determined by the driving frequency of the AOM.   

%
Considering the fact that the driving frequency of the commercially available AOM is up to hundreds MHz and can be extended to GHz range in principle \cite{Wan2022}, we think the frequency tuning ability of the ABI can be further extended to GHz range. 
%
%
Moreover, other than squeezed vacuum state, the frequency tuning scheme can also be applied to a variety of other quantum optical states such as single photon state or entangled state, and can serve as a handy tool to realize coherent interaction between the quantum optical states with optical frequency difference. 

The paper is organized as follows. Section II briefly introduces the experimental principles, including the frequency tuning by the AOM based bi-frequency interferometer and the measurement scheme for frequency tuned squeezed vacuum state. 
Section III demonstrates the experimental setup. 
Section IV presents the experimental procedures and results. We conclude the papaer in Section V with discussion and summary.

\section{Experimental principle}
Fig.1 (a) shows the schematic diagram of the ABI \cite{2022arXiv221000406L}, where the AOMs, based on the photon-phonon interaction between the incident light and the medium with a time-varying stratified refractive index, are used to replace the beam splitters in a conventional Mach-Zehnder interferometer. 
%
%
For a single AOM, say AOM1 in Fig. 1(a), when the input fields $\hat a$ and $\hat b$ with optical frequency $\omega+\Omega$ and $\omega$ are aligned along the -1$^{st}$ and +1$^{st}$ order diffraction direction, the output fileds $\hat c$ and $\hat d$ is related to the inputs field $\hat a$ and $\hat b$ (satisfying the standard bosonic commutation relation $[\hat a, \hat a^\dagger]=[\hat b, \hat b^\dagger]=1$) through
%
\begin{eqnarray}
	\label{AOM_1}
	{\hat c} &=& t  \hat b_{\omega+\Omega} + r  \hat a_{\omega} \nonumber\\
	{\hat d} &=& t  \hat a_{\omega} - r  \hat b_{\omega+\Omega},
\end{eqnarray}
where the subscripts of the operators $\omega$ and $\omega+\Omega$ denote the optical frequencies, $t$ and $r$ satisfying $|t|^2+|r|^2=1$ represent the effective transmission and reflection coefficients related to the strength of the Bragg diffraction process.
Note Eq. (\ref{AOM_1}) is the same as the equation for the quantum state frequency conversion through parametric processes \cite{Kumar1990OL} except that here the frequency change quantity $\Omega$ depends on the frequency of phonons.
Currently, the diffraction efficiency of commercial available AOMs still can not approach to ideal due to high order diffraction. To realize frequency conversion near ideal efficiency, we exploit the interferometric enhanced Bragg diffraction effect by introducing another AOM (AOM$_2$) to form the bi-frequency interferometer depicted in Fig. 1(a).
Assuming the power splitting ratio of both AOMs are set to $ |t|^2 = |r|^2 = 50\%$, the output optical field of AOM$_2$ denoted by $\hat e$ and $\hat f$ reads
 \begin{eqnarray}
	\label{ABI_as_splitter}
	&{\hat e}_{\omega+\Omega} = \frac{e^{i\varphi}-1}{2} {\hat a}_{\omega+\Omega} - \frac{e^{i\varphi}+1}{2} {\hat b}_{\omega+\Omega} \nonumber\\
	&{\hat f}_{\omega} =  \frac{e^{i\varphi}+1}{2} {\hat a}_{\omega} + \frac{e^{i\varphi}-1}{2} {\hat b}_{\omega},  
\end{eqnarray}
where $\varphi$ is the phase difference between two arms of the ABI. 
Eq. (\ref{ABI_as_splitter}) clearly illustrates that interferometer formed by two AOMs functions as a beam splitter for two input fields with frequency $\omega$ and $\omega+\Omega$, respectively, and the splitting ratio depends on $\varphi$.
For $\varphi$=0, the quantum state $|\Psi\rangle_\omega$ at $\omega$ in port $\hat b$ is completely transferred to port $\hat e$ with frequency shifted to $\omega+\Omega$. 

\begin{figure*}[ht]
	\includegraphics[width=0.95\textwidth]{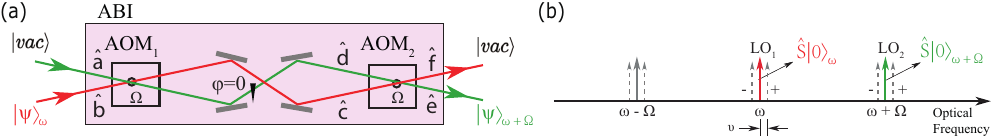}
	\caption{\label{fig1}
		(a) The schematic diagram of high efficiency frequency tuning device formed by two AOMs. 
		(b) Illustration of the frequency domain homodyne detection scheme for measuring the squeezed vacuum state with frequency $\omega$ and $\omega+\Omega$, respectively.
	}
\end{figure*}

We note Eq. (1-2) assume the ABI has perfect optical efficiency and mode matching. In practical, the systematic efficiency of the ABI functions as a frequency tuning device can be written as \cite{2022arXiv221000406L}
\begin{equation}
	\label{eta_ABI}
	\eta_{ABI}= \frac{\zeta}{2}( 1 + V ),
\end{equation}
where $\zeta$ is the efficiency of optical elements for individual arm (assuming the loss of the two arms is the same), and 
$V=(I_M-I_m)/(I_M+I_m)$ is the visibility of the ABI determined by the mode matching ($I_M$ and $I_m$ are the maximum and the minimum intensity measured at either output port when varying $\varphi$).
%
For $\eta=1$, when the two inputs of ABI are in vacuum state $|vac\rangle$ and squeezed vacuum state ($\hat S |0\rangle_{\omega}$), respectively, port $\hat e$ gives out $\hat S |0 \rangle_{\omega+\Omega}$, which is the same as the input state except the optical frequency is tuned by $\Omega$, and port $\hat f$ gives out a vacuum state $ |vac\rangle$. For non-ideal systematic efficiency $\eta_{ABI}$, port $\hat e$ gives out a classical mixture of $\hat S |0 \rangle_{\omega+\Omega}$ and $ |vac\rangle$ leading to the reduction of squeezing rate.

The squeezed state is measured by homodyne detection (HD) in frequency domain.
For a squeezed state generated from an optical parametric oscillator, the maximum squeezing resident in the degenerate frequency of the OPO (noted by $\omega$).
However, technical noise prevent one to measure exactly at the degenerate, and a side band measurement at a detuning frequency within the bandwidth of the OPO is commonly preformed to characterize a squeezed state \cite{PhysRevLett.117.110801}. 
This kind of measurement is essentially a two-mode squeezing measurement \cite{{PhysRevA.71.041802},{PhysRevA.92.012110},{PhysRevLett.111.200402}}. 
The HD with local oscillator (LO) centering at $\omega_L$ measures a pair of side band modes with symmetric frequency detuning to the local oscillator field.
The normalized noise power of the homodyne detection $S_{\mathrm{HD}}$ is related to the side band modes by \cite{{PhysRevLett.111.200402},{PhysRevA.88.052113}, {PhysRevLett.95.243603}}
\begin{eqnarray}
	\label{PowerSpecHD}
 	S_{\mathrm{HD}}(\theta) = & \frac{\cos ^2 \theta}{2}  \big( \langle \Delta^2 \hat{X}_{\omega_{L}}^{(+)} \rangle + \langle \Delta^2 \hat{P}_{\omega_{L}}^{(-)} \rangle \big)  + \frac{\sin ^2 \theta}{2}  \big( \langle\Delta^2 \hat{X}_{\omega_{L}}^{(-)}\rangle+\langle\Delta^2 \hat{P}_{\omega_{L}}^{(+)}\rangle \big)
\end{eqnarray}
with
\begin{eqnarray}
	\label{sideband}
	\hat{X}_{ \omega_{L}}^{(\pm)} = & \frac{1}{\sqrt{2}}\left(\hat{X}_{\omega_{L}+\nu} \pm \hat{X}_{\omega_{L}-\nu}\right) \nonumber\\
	\hat{P}_{ \omega_{L}}^{(\pm)} = & \frac{1}{\sqrt{2}}\left(\hat{P}_{\omega_{L}+\nu} \pm \hat{P}_{\omega_{L}-\nu}\right),
\end{eqnarray}
where $\nu$ is the electronic analyzing frequency, $\theta$ is the phase between the LO and the input state, and  
%
$\hat{X}_{\omega_{L}\pm\nu}$ and $\hat{P}_{\omega_{L}\pm\nu}$ are the quadrature amplitude and phase operators defined as $\hat{X}_{\omega_{L}\pm\nu} = (1/\sqrt{2}) (\hat{a}_{\omega_{L}\pm\nu} + \hat{a}^\dagger_{\omega_{L}\pm\nu})$ and $\hat{P}_{\omega_{L}\pm\nu} = (-i/\sqrt{2}) (\hat{a}_{\omega_{L}\pm\nu} - \hat{a}^\dagger_{\omega_{L}\pm\nu})$ with $\hat{a}_{\omega_{L}\pm\nu}$ and $\hat{a}^\dagger_{\omega_{L}\pm\nu}$ being the annihilation and the creation operator satisfying the canonical commutation relation $[\hat{a}_{\omega_{L}\pm\nu}, \hat{a}^\dagger_{\omega_{L}\pm\nu}]=1$.
At the limit of $\nu \to 0$, the noise power given by Eq. (\ref{PowerSpecHD}) becomes the fluctuation of the quadrature amplitude $\hat X_{\omega_L}$ and quadrature phase $\hat P_{\omega_L}$ of the light field  for $\theta = 0$ and $\theta = \pi/2$, respectively.
%

%
In the process of HD measurement, the frequency mode match between the squeezed  state and local oscillator plays an important role. Fig. 1 (b) illustrates the frequency relation for measuring the squeezed states $ \hat S |0 \rangle_\omega$ and  $ \hat S |0 \rangle_{\omega+\Omega}$ when the LO frequency of homodyne detection (HD) is $\omega$ and $\omega+\Omega$, respectively. For the sake of clarity, the cases of $\omega_L=\omega$ and $\omega_L=\omega+\Omega$ are labeled as LO$_1$ and LO$_2$.
In the experiment of calibrating $ \hat S |0 \rangle_\omega$, HD using LO$_1$ is performed at electronic analyzing frequency $\nu$.
However, to measure $ \hat S |0 \rangle_{\omega+\Omega}$, both LO$_1$ and LO$_2$ are used. 
By observing the beat signal between $ \hat S |0 \rangle_{\omega+\Omega}$ and LO$_1$, we verify the frequency tuning effect. In this case, the side band pairs for a given $\nu$ are shifted from $\omega + \nu$ and $\omega - \nu $ to $\omega+ \Omega + \nu$ and $\omega+ \Omega - \nu $, which is not symmetric to LO$_1$ and squeezing rate can not be directly extracted from the measurement.
To calibrate the squeezing rate of $ \hat S |0 \rangle_{\omega+\Omega}$, we need to use LO$_2$ and analyze at the same electronic frequency $\nu$ as it is for calibrating $ \hat S |0 \rangle_{\omega}$.
%


\section{Experimental setup and phase locking scheme}

Our experimental setup is shown in Fig. 2 (a). 
A continuous wave laser with center wavelength of 1550.12 nm and spectrum line-width of 750 Hz is frequency doubled by a second harmonic generator (SHG). The output second harmonic field with optical frequency $2\omega$ is used as the pump for a bow-tie configured OPO using periodically poled potassium titanyl phosphate (PPKTP) crystal as the gain media \cite{xguo2020}. 
The squeezed vacuum state $ \hat S |0 \rangle_{\omega}$ is generated from the OPO below threshold.
Passing the squeezed state $ \hat S |0 \rangle_{\omega}$ through ABI, which functions as a frequency tuning device, the optical frequency of squeezed state is shifted to $\omega+\Omega$. The squeezed state is measured by using a homodyne detection (HD) system, whose LO frequency is either LO$_1$ or LO$_2$. The output of the HD system is recorded by a 14-bits oscilloscope (NI PXIe-5170R).
For the sake of convience, the functional models of the setup, including OPO, ABI and HD system (see Fig. 2(a)), are coupled by fiber collimators (C$_1$ to C$_4$) and optically connected by single mode fibers (represented by yellow bold lines in Fig. 2).
The quantum efficiency of the fiber based setup has not been optimized to reach that of a free space implementation yet, but the flexibility of the setup is greatly enhanced.  

\begin{figure*}[ht]
	\includegraphics[width=0.9\textwidth]{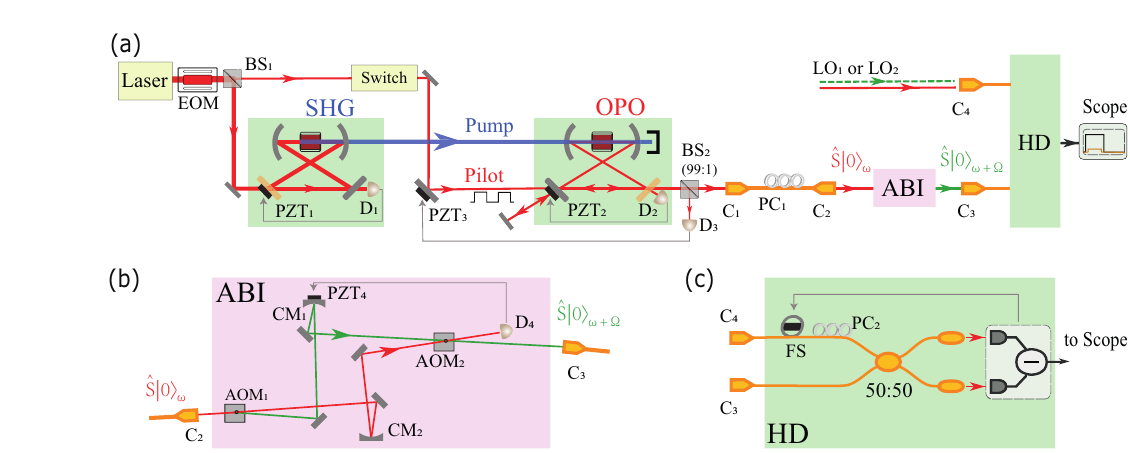}
	\caption
	{
		(a) The experimental setup. (b) and (c) plot the details of frequency tuning device and homodyne detection system, respectively. EOM, electro-optical modulator; SHG, second harmonic generator; OPO, optical parametric oscillator; ABI, acousto-optic modulator (AOM) based bi-frequency interferometer; HD, homodyne detection; LO, local oscillator for HD; BS, beam-splitter; C, fiber collimator; PC, polarization controller; PZT, piezoelectric lead zirconate titanate based phase shifter; D, photo-diode for phase locking;
		CM, concave mirror;
		FS, fiber stretcher. 
	}
\end{figure*}

Fig.2 (b) shows the details of the ABI formed by two AOMs (ISOMET M1205).
The squeezed vacuum state $ \hat S |0 \rangle_{\omega}$  is coupled to the +1$^{st}$ diffraction mode ($\hat b$ in Fig. \ref{fig1}(a)) of AOM$_1$ through the fiber collimator (C$_2$), and 
splitted into a frequency up-shifted arm (green beam with the frequency of $\omega+\Omega$) and a non-shifted arm (red beam with the frequency of $\omega$) by AOM$_1$. 
Then the two arms are coherently combined by AOM$_2$. 
For each arm between two AOMs, two plane mirrors and a concave mirror (CM) with 300 mm radius of curvature are used to ensure the required mode matching is optimized.
In our experiment, two radio frequency signals with frequency of $\Omega=80$ MHz drive the two AOMs to set the diffraction efficiency for each AOM around 50$\%$.
To control the phase $\varphi$ between two arms, a piezo phase shifter (PZT$_4$) is mounted on the concave mirror of the frequency shifted arm.
$\hat{S}|0\rangle_\omega$ is converted into $\hat{S}|0\rangle_{\omega+\Omega}$ and is coupled to the input of HD system through a fiber collimator (C$_3$). 


Fig.2 (c) shows the configuration of the HD system, which is implemented by using a fiber 50:50 beam splitter and a balanced detector. 
LO$_1$ and LO$_2$ with optical frequency $\omega$ and $\omega+\Omega$ are obtained by sending a small fraction of the laser through an AOM (not shown in Fig.2) and collecting the 0$^{th}$ and the +1$^{st}$ diffracted light, respectively. During the experiment, either LO$_1$ or LO$_2$ is sent into the fiber coupled HD. 
The squeezed vacuum state are injected into the HD system through C$_3$. The light from two output ports of the 50:50 is sent onto a pair of high efficiency photodiodes.  
The polarization controller (PC$_2$) is used to optimize the polarization mode matching, and the fiber stretcher (FS) is used to control the phase $\theta$ between the LO and the squeezed state.

\begin{figure*}[ht]
	\includegraphics[width=0.55\textwidth]{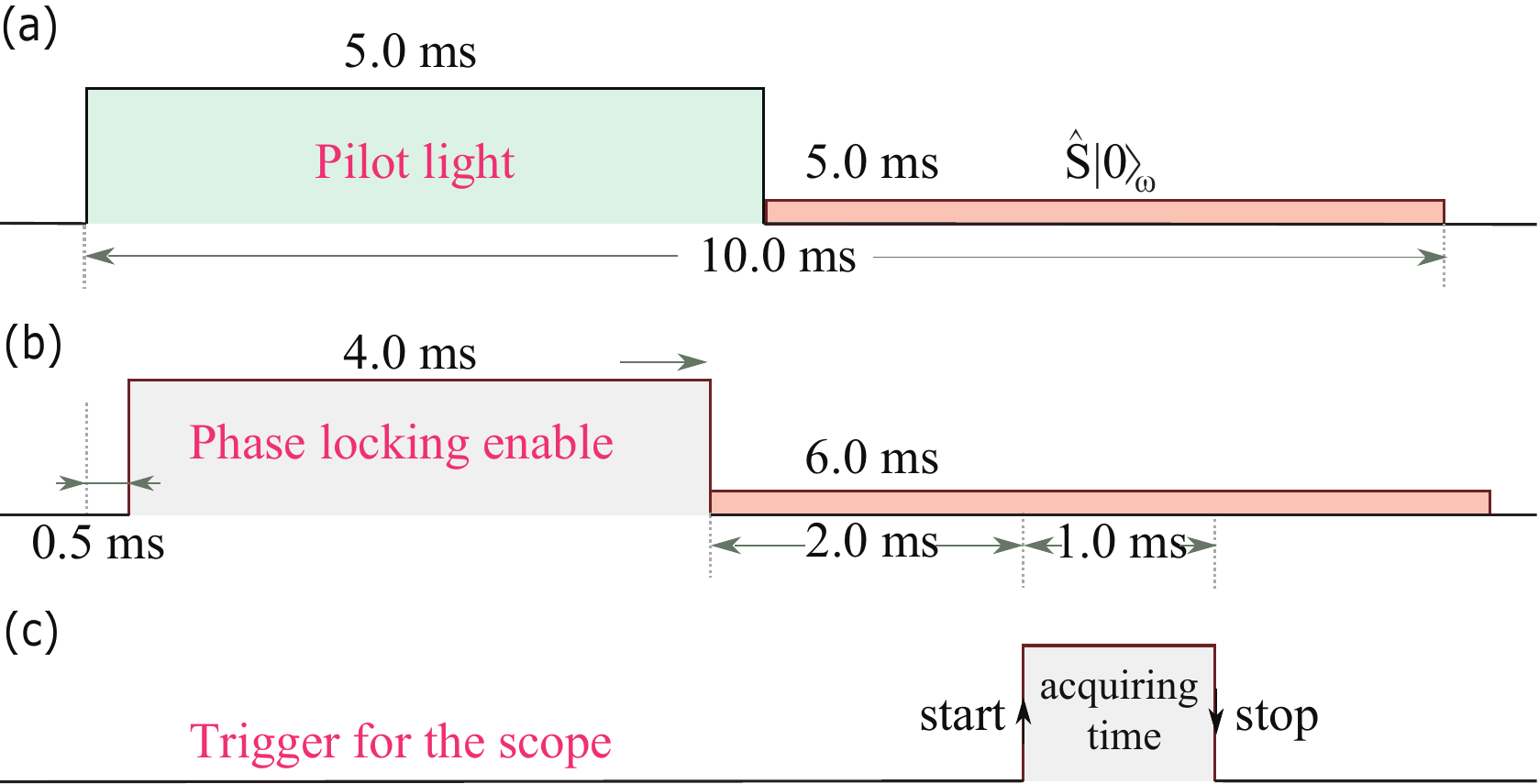}
	\caption
	{
		The time controlling sequence of the phase locking and measurement data acquiring process. (a) The light output from the OPO. (b) The phase locking enable signal applied on the locking process of the OPO cavity resonance, the parametric gain, the ABI and the HD system. (c) The trigger signal for the scope in the data acquisition process. 
	}
\end{figure*}

In the process of measuring $\hat S |0\rangle_{\omega}$, the OPO is operated at resonant condition, the phase between LO and squeezing is locked at $\theta=0$. When the frequency of squeezing is shifted to $\omega+\Omega$, in addition to above conditions, the ABI need to be locked at $\varphi=0$.
To ensure the stable operation of experimental setup, five sets of phase locking loops are implemented by using field programmable gate arrays (FPGAs, STEMlab125-14) based feedback systems. 
Except for locking the cavity resonance of SHG, all the other phase locking loops work in period locking mode. 
A pilot light with 50$\%$ duty cycle and 100 Hz generated by an optical switch is injected into the OPO and used as the phase reference of the period locking. 
When the pilot light is on, the feedback loops of OPO, ABI and HD are active. When the pilot light is off, the squeezed vacuum state is generated from OPO and the data acquisition system of HD starts to take data. We assume that locking condition established by the feedback loops will not drift that much when pilot light is temporally off for the short period of 5 ms. 

The details on the time controlling sequence of the period phase locking scheme is illustrated in Fig. 3 (a-c). 
%
%
%
%
%
Five gray traces with each one started from a detector (D$_i$, i=1,2,3,4 or the balanced detector in HD) and ended by an arrow pointing to a phase variable components (PZT or FS) in Fig. 2 represent our phase control system.
%
The cavity resonance frequency of SHG is locked to the optical frequency of the laser $\omega$ by the feedback loop formed by D$_1$  and PZT$_1$, and similar locking for OPO is carried out by the feedback loop from D$_2$ to PZT$_2$.
Both cavities are locked by using Pound-Drever-Hall (PDH) method \cite{PDH} with a phase modulation signal at 13 MHz generated by a phase modulator (EOM). 
The feedback loop formed by D$_3$ and PZT$_3$ stabilize the phase between the pump and the pilot light so that the pilot light is maximally de-amplified. The input light to D$_3$ is obtained by using a beam splitter with 99:1 splitting ratio (BS$_2$).
The feedback loop from D$_4$ to PZT$_4$ stabilizes the phase $\varphi$ between the two arms of the ABI. To realize the locking of $\varphi=0$, a small phase modulation of 200kHz is added on the 80 MHz radio frequency signal that drives AOM$_1$ and digital demodulation is implemented in the FPGA \cite{2022arXiv221000406L}.
The HD phase $\theta$ is stabilized by the feedback loop inside HD system, where the HD output signal is sent to a fiber stretcher (FS) that can change the phase of the LO. A phase dithering signal generated from the FPGA is also applied onto the FS, which enables the HD phase $\theta$ to be locked to arbitrary value.

When the function models of experimental setup, including OPO, ABI and HD, are properly locked, the terms $\langle\Delta^2 \hat{X}^{(\pm)}_{\omega_{L}}\rangle+\langle\Delta^2 \hat{P}^{(\mp)}_{\omega_{L}}\rangle$ in Eq. (\ref{PowerSpecHD}) can be expressed as \cite{Collett1984PRA}
\begin{eqnarray}
	\label{sq_asq}
	\langle\Delta^2 \hat{X}^{(\pm)}_{\omega_{L}}\rangle+\langle\Delta^2 \hat{P}^{(\mp)}_{\omega_{L}}\rangle  =  1 \mp \eta\frac{4\sqrt{P/P_{th}}}{(1 \pm \sqrt{P/P_{th}} )^2+4(\nu / \nu_0)^2},
\end{eqnarray} 
where $\eta$ is the overall detection efficency, $P$ is the pump power, and $\nu$ is the electronic analyzing frequency. $P_{th} = 980$ mW is the threshold power of the OPO, and $\nu_0 = 15.6$ MHz is the bandwidth of the OPO.
%

\section{Experimental procedures and results}%
We first calibrate the squeezed vacuum state $\hat{S}|0\rangle_\omega$ by removing the frequency tuning device (ABI) and coupling the transmittance light of BS$_2$ into HD system (see Fig. 2 (a)).
LO$_1$ with optical frequency $\omega$ is used as local oscillator of HD.
The squeezed quadrature 
$\langle\Delta^2 \hat{X}^{(+)}_{\omega}\rangle+\langle\Delta^2 \hat{P}^{(-)}_{\omega}\rangle$ 
and anti-squeezed quadrature 
$\langle\Delta^2 \hat{X}^{(-)}_{\omega}\rangle+\langle\Delta^2 \hat{P}^{(+)}_{\omega}\rangle$ 
can be measured by locking the HD phase to $\theta=0$ and $\theta = \pi/2$, respectively (see Eqs. (\ref{PowerSpecHD}-\ref{sq_asq})).
%
%
Fig. 4(a) plots a set of typical results of the measured power spectrum, which is obtained when the pump power for the OPO is $P=450$ mW and the LO power is 500 $\mu$W. The scope records 50K successive voltage values with a sampling rate of 50 MHz. 
For each trace in Fig. 4 (a), 500 measurement rounds are implemented and the averaging results of the fast Fourier transform obtained from the recorded voltage values are used to estimate the power spectrum.
To obtain the squeezing and anti-squeezing at electronic analyzing frequency $\nu = 1.55$ MHz, we integrate the power spectrum centered at 1.55 MHz with an integration bandwidth of 0.1 MHz. 
We also record the shot noise level (SNL) and electronic noise of HD, respectively. The former is obtained by blocking the signal input so that HD measures the noise of vacuum; the latter is obtained by blocking both the signal input and LO of HD.
After correcting the influence of electronic noise, the squeezing and the anti-squeezing are about 3.02 $\pm$ 0.02 dB below SNL (blue trace) and 11.64 $\pm$ 0.02 dB above SNL, respectively.
%

\begin{figure*}[ht]
	\includegraphics[width=0.65\textwidth]{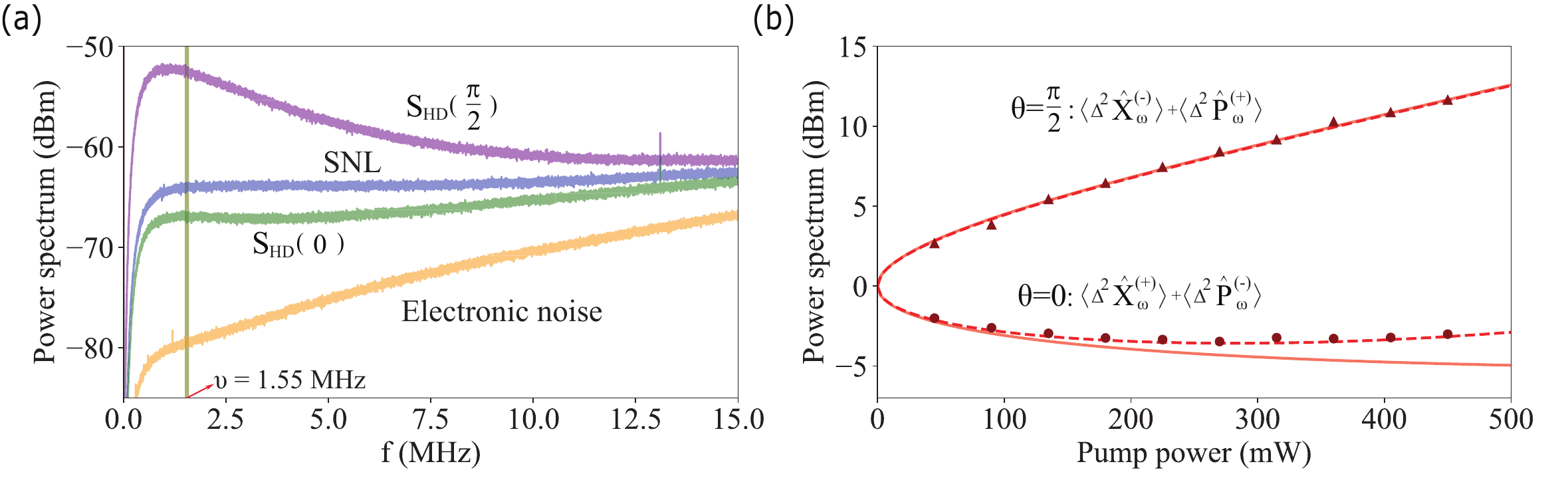}
	\caption{
		Squeezed vacuum state $\hat S |0\rangle_\omega$ measured by homodyne detection .
		(a) The power spectrum for squeezing ($S_{\mathrm{HD}}$(0)) and anti-squeezing ($S_{\mathrm{HD}}$($\pi$/2)) of $\hat S |0\rangle_\omega$, when the pump power is $P= 450 $ mW. 
		%
		%
		(b) The measured squeezing and anti-squeezing rate of $\hat S |0\rangle_\omega$ at different pump power. In the measurement, the electronic analyzing frequency is $\nu = 1.55$ MHz.
		The solid curves in (b) are the theoretical predictions of the noise power normalized to SNL by Eq. (\ref{PowerSpecHD}-\ref{sq_asq}) for ${\theta}= 0$ and ${\theta}= \pi/2$, respectively. The dashed curves are the theoretical predictions when the imperfect phase locking is modeled by a phase offset of ${\Delta\theta}= 6^\circ$.
	}
\end{figure*}

Fig. 4 (b) shows the measurement results of squeezing when the pump power is varied.
As the pump power increases, the noise of the squeezing quadrature 
first decreases and then increase but the noise of the anti-squeezing 
keep increasing. 
A maximum squeezing of $\hat{S}|0\rangle_\omega$ is -3.47 $\pm$ 0.02 dB, which is obtained at the pump power of $P = 270 $mW. 
The red solid curves in Fig. 4 (b) are obtained by substituting the experimental parameters into Eq. (\ref{PowerSpecHD}-\ref{sq_asq}), in which the total detection efficiency is $\eta = 70.8\%$ (including the escape efficiency of the OPO cavity $93.4 \%$, the coupling efficiency between OPO and HD $ 85.4\%$, the systematic detection efficiency of HD $88.8\%$).
One sees the anti-squeezing data points ($\theta = \pi/2$) well fit with the theoretical curve but the squeezing data points ($\theta = 0$) slightly deviate from the solid curve. We think this is because the phase locking system is imperfect, and the measurement of squeezing quadrature is more sensitive to the LO phase than that of anti-squeezing quadrature. 
This imperfection on phase locking can be modeled by a small angle offset \cite{oe-14-15-6930}, and the dashed lines in Fig. 4 (b) is obtained by introducing an offset angle of $\Delta \theta = 6 ^{\circ}$ to Eq. (3) which fits the data better.
The comparison of experimental data and theoretical prediction in Fig. 4(b) indicates that squeezing rate is currently limited by the transmission and detection efficiency, and the quality of phase locking.

Then we sent the squeezed vacuum state $\hat{S}|0\rangle_\omega$ through the ABI to obtain the frequency tuned state $\hat{S}|0\rangle_{\omega+\Omega}$.
To show the frequency shifting property, LO with frequency centering at $\omega$ (LO$_1$) is used to measure the beating noise signal.
Additionally, two modifications on the state detection process are made to efficiently measure the beating noise signal at 80 MHz: (1) We replace the HD setup in the Fig. 2 with a fast response one \cite{Zhao:23}, which has a bandwidth of about 100 MHz, and the LO power is increased to 2 mW to ensure the SNL of HD is still well above electronic noise; (2) The sampling rate of the scope is increased to 250 MHz to resolve the signal at around 80 MHz.
The measurement of noise power spectrum with local phase $\theta = \pi/2$ is shown in Fig.5 (a), where the OPO is running at the same condition for obtaining Fig. 4 (a). The result obviously shows the squeezed vacuum state is frequency shifted by $\Omega= 80$ MHz, and the noise is 4.34 $\pm$ 0.01 dB above SNL at the analyzing frequency $\Omega+\nu$=81.55 MHz or $\Omega-\nu$=78.45 MHz. 
We also change the LO phase and observe no noticeable difference, which proves the result is phase insensitive and always larger than the SNL. 
This property is due to the HD system measures a mixture of a single mode of the side band mode pair at frequency $\omega+\Omega\mp\nu$ and a vacuum state at frequency $\omega-\Omega\pm\nu$. Note that a peak at 80 MHz in Fig. 5 (a) can be observed. We think is caused by the electronic coupling of the high power 80 MHz radio frequency signal that drives the AOMs.
For the convenience of further analysis (see Eqs. (\ref{sq_by_r}-\ref{th_by_r})), we evaluate the overall detection of this experiment, which is $43.9\%$ (including the propagation efficiency before the input port the ABI $71.3\%$, the systematic efficiency of the ABI $91\%$ \cite{2022arXiv221000406L}, the coupling efficiency between the ABI and the HD $ 84.1\%$, and the systematic detection efficiency of the fast HD $80.6\%$).

\begin{figure}[ht]
	\includegraphics[width=0.95\textwidth]{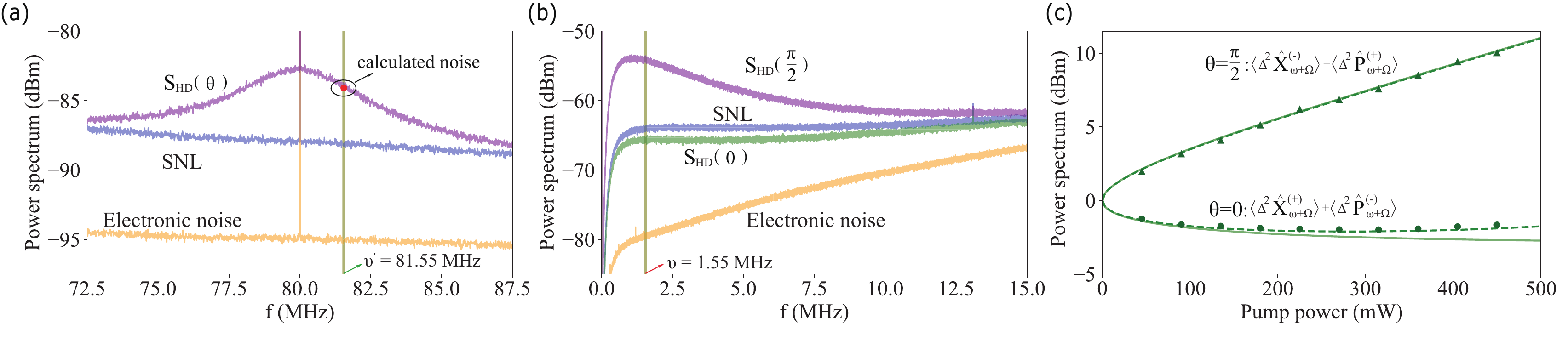}
	\caption{Results for the frequency tuned squeezed vacuum state $\hat S |0\rangle_{\omega+\Omega}$.
		(a) The power spectrum of the squeezed state $\hat S |0\rangle_{\omega+\Omega}$ when measured by LO$_1$ with the frequency of $\omega_L=\omega$. (b) The power spectrum of the squeezed state $\hat S |0\rangle_{\omega+\Omega}$ when measured by LO$_2$ with the frequency $\omega_L=\omega+\Omega$. Results in both (a) and (b) are obtained at the same pump power of $P$ = 450 mW.
		%
		%
		(c) The squeezing and anti-squeezing rate of $\hat S |0\rangle_{\omega+\Omega}$ at analyzing frequency v=1.55 MHz in different pump power when $\omega_L=\omega+\Omega$.
		The solid curves in (c) are the theoretical predictions of the noise power normalized to SNL by Eq. (\ref{PowerSpecHD}-\ref{sq_asq}) for ${\theta}= 0$ and ${\theta}= \pi/2$, respectively, while the dashed curves are the theoretical predictions when including the imperfection of phase locking by a phase offset of ${\Delta\theta}= 6^\circ$. 
		}

\end{figure}

Having shown the frequency shifted property, we measure the squeezing rate of $\hat{S}|0\rangle_{\omega+\Omega}$ using LO with frequency centering at $\omega+\Omega$ (LO$_2$), and the HD system is switched back to the one used in obtaining the data in Fig. 4. 
%
A typical measurement is shown Fig. 5 (b), where the operation condition of OPO is still the same as that for obtaining data in Fig. 4(a). 
With similar data analysis approach, the squeezing and anti-squeezing at $\nu=1.55$ MHz is estimated to be -1.66 $\pm$ 0.02 dB and 10.02 $\pm$ 0.02 dB (after correcting the electronic noise), respectively. Compared to the result in Fig. 4 (a), both squeezing and anti-squeezing results in Fig. 5(b) are more close to the SNL.
%
The squeezing and the anti-squeezing of $\hat{S}|0\rangle_{\omega+\Omega}$ is also measured in different pump power. The result is shown in Fig. 5(c), where a maximum squeezing of $\hat{S}|0\rangle_{\omega+\Omega}$ is -1.98 $\pm$ 0.02 dB at the pump power of 270 mW. 
Comparing Fig. 5(c) with Fig. 4(b), one sees that the variation trend of the squeezing and anti-squeezing is the same as the squeezed vacuum state $\hat{S}|0\rangle_{\omega}$, but the squeezing rate is lower. 
We argue this squeezing rate reduction is caused by the extra loss. The overall efficiency in this case is $48.3 \%$ (we note the systematic efficiency of the HD system in this measurement is $88.8\%$, which is higher than the fast HD system used to obtain Fig. 5 (a)). 
The theoretical curves for no phase offset (solid line) and that for an offset angle of $\Delta \theta = 6 ^{\circ}$ (dashed line) are also plotted in Fig. 5 (c). The consistency between the experimental data and the theoretical curves is similar to Fig 4 (b), which indicates the reduction of squeezing rate is due to optical loss, and other extra noise introduced by the frequency tuning device is negligibly small.

Finally, it is interesting to compare the result in Fig. 5 (a) to the anti-squeezing power spectrum in Fig. 5 (b), which are the same state but measured by two different HD systems. 
For a given electronic analyzing frequency $\nu$ and an overall efficiency $\eta$, one can introduce an effective squeezing parameter $r$. Fig. 5 (b) is obtained by using a LO$_2$ with optical frequency $\omega+\Omega$, so Eq. (4) can be rewritten as \cite{Gerry_Knight_2004}
\begin{eqnarray}%
	\label{sq_by_r}
	\langle\Delta^2 \hat{X}^{(\pm)}_{\omega+\Omega}\rangle+\langle\Delta^2 \hat{P}^{(\mp)}_{\omega+\Omega}\rangle = \eta(\sinh r \mp \cosh r)^2+(1-\eta), 
\end{eqnarray} 
For the results in Fig. 5 (b), the effective squeezing parameter $r$ for $\nu=1.55$ MHz can be estimated by using the measured result and substituting $\eta=48.3\%$ into Eq. (7), which yields $r = 1.49$.

For the result in Fig. 5 (a), LO$_1$ with optical frequency of $\omega$ is used and the side band pairs with quantum correlation is not symmetric to LO$_1$. Therefore, the homodyne detection measures a mixture of a vacuum state at frequency $\omega-\Omega\mp\nu$ and a single mode of the side band mode pair at frequency $\omega+\Omega\pm\nu$. 
Considering the vacuum state of the side band pairs will introduce 50$\%$ loss to the output noise, the output noise at $\Omega \pm \nu = 80 \pm 1.55$ MHz can be expressed as \cite{Gerry_Knight_2004}
\begin{equation}
	\label{th_by_r}
	\langle\Delta^2 \hat{X}^{(\pm)}_{\omega}\rangle+\langle\Delta^2 \hat{P}^{(\mp)}_{\omega}\rangle = \frac{\eta}{2}({\sinh^2 r} + {\cosh^2 r})+\frac{1-\eta}{2}.
\end{equation}
We calculate the noise $\langle\Delta^2 \hat{X}^{(\pm)}_{\omega}\rangle+\langle\Delta^2 \hat{P}^{(\mp)}_{\omega}\rangle$ at $\nu' =\Omega + \nu  = 81.55$ MHz by substituting $r=1.49$ and $\eta = 43.9 \% $ into Eq. (7). The calculated result is plotted with a red circle dot in Fig. 5(a), indicating that measurement results agree with the theoretical expectation.

\section{Conclusion}
In summary, we demonstrate a high efficiency frequency tuning scheme for continuous variable quantum state. 
Using the frequency tunning device formed by an acousto-optics modulator based bi-frequency interferometer, the frequency of squeezed state at 1550.12 nm is up shifted by 80 MHz. 
The frequency tuning property is proven by directly measuring the beating noise around 80 MHz, and the noise power spectra of the squeezed state before and after frequency tuning are further analyzed by using LO with different frequencies. 
The squeezing rate after frequency tuning is mainly confined by the fiber coupling loss and the system efficiency of the ABI, and the extra noise introduce by the frequency tuning process is negligible. The systematic efficiency of the ABI is up to 91$\%$ and can be further improved in principle. With this efficiency, a squeezed vacuum state of -13 dB squeezing rate \cite{Schonbeck2018OL} will still preserve -9 dB,illustrating that the setup can be used in quantum information technology.     
Although the amount of frequency shift in our experiment is 80 MHz, the tuning range is determined by the performance of AOMs,
%
which can in principle be optimized to be up to GHz range. 
These properties makes our scheme not only useful for coherently interaction among quantum optical states, but also can be applied to the other type of quantum systems. For example, the quantum frequency tuning scheme can be used to change the frequency of single photon to match the spectrum line of a cold atom \cite{TiancaiZhang2023PRL}.

In addition, our result implies that the interferometric enhancement method can be extended to that of the quantum frequency conversion realized by parametric process \cite{Huang1992PRL, Clark2013OL, Wang2023}. 
By using cascaded parametric processes, the power required for the pump of the parametric process can be greatly reduced. Therefore, the other nonlinear effects serving as the noise source for the frequency conversion process might be mitigated.

%

\begin{acknowledgments}
This work was supported in part by National Natural Science Foundation of China (Grants No.12004279 and
12074283).
\end{acknowledgments}


%

\end{document}